\documentclass[12pt,preprint]{aastex}
\usepackage {graphicx}
\usepackage{aastexug}

\shorttitle{Screw instability of the magnetic field}
\shortauthors{D. X. Wang et al.}

\begin{document}

\title{Screw instability of the magnetic field connecting a
rotating black hole with its surrounding disk}
\author{Ding-Xiong Wang\altaffilmark{1},  Ren-Yi Ma  and  Wei-Hua Lei}
\affil{Department of Physics, Huazhong University of Science and
Technology,Wuhan, 430074, P. R. China} \and
\author{Guo-Zheng Yao}
\affil{Department of Physics, Beijing Normal University, Beijing,
100875, P. R. China}

\altaffiltext{1}{Send offprint requests to: D. X. Wang
(dxwang@hust.edu.cn)}

\begin{abstract}

Screw instability of the magnetic field connecting a rotating black hole
(BH) with its surrounding disk is discussed based on the model of the
coexistence of the Blandford-Znajek (BZ) process and the magnetic coupling
(MC) process (CEBZMC). A criterion for the screw instability with the state
of CEBZMC is derived based on the calculations of the poloidal and toroidal
components of the magnetic field on the disk. It is shown by the criterion
that the screw instability will occur, if the BH spin and the power-law
index for the variation of the magnetic field on the disk are greater than
some critical values. It turns out that the instability occurs outside some
critical radii on the disk. It is argued that the state of CEBZMC always
accompanies the screw instability. In addtition, we show that the screw
instability contributes only a small fraction of magnetic extraction of
energy from a rotating BH.
\end{abstract}



\section{INTRODUCTION}

It is well known that the magnetic field configurations with both
poloidal and toroidal components can be screw instable (Kadomtsev
1966; Bateman 1978). According to the Kruskal-Shafranov criterion
(Kadomtsev 1966), the screw instability will occur, if the
toroidal magnetic field becomes so strong that the magnetic field
line turns around itself about once. Recently some authors
discussed the screw instability in black hole (BH) magnetosphere.
Gruzinov (1999) argued that a Kerr BH, being connected with a disk
by a bunch of closed field lines, can flare quasi-periodically,
and this process can be regarded as a new mechanism for extracting
rotational energy from the BH. Li (2000a) discussed the screw
instability of the magnetic field in the Blandford-Znajek (BZ)
process (Blandford {\&} Znajek 1977), which results in a stringent
upper bound to the BZ power. Tomimatsu et el. (2001) studied the
condition of the screw instability in the framework of the
variation principle, and argued that the field-line rotation has a
stabilizing effect against the screw instability.

Recently much attention has been paid on the magnetic coupling
(MC) process, where energy and angular momentum are extracted from
a fast-rotating BH to its surrounding disk by virtue of closed
magnetic field lines (Blandford 1999; Li 2000b, 2002a; Wang, Xiao
{\&} Lei 2002, hereafter WXL; Wang, Lei {\&} Ma 2003, hereafter
WLM). The MC process can be used to explain a very steep
emissivity in the inner region of the disk, which is found by the
recent \textit{XMM-Newton} observation of the nearby bright
Seyfert 1 galaxy MCG-6-30-15 (Wilms et al. 2001; Li 2002b; WLM).
Compared with the BZ process, the disk load in the MC process is
much better understood than the remote load. Very recently we
discussed the condition for the coexistence of the BZ and MC
processes (CEBZMC), and some electromagnetic quantities, such as
electric field, magnetic field and electric current in the BH
magnetosphere, are determined (Wang et al. 2003, hereafter WMLY).

The above works motivate us to discuss the screw instability in
CEBZMC based on the Kruskal-Shafranov criterion and our
calculations of the poloidal and toroidal components of the
magnetic field. This paper is organized as follows. In \S 2 we
give a brief description on our model of CEBZMC. In \S 3 the
criterion of the screw instability is derived based on the
Kruskal-Shafranov criterion and our calculations of the poloidal
and toroidal components of the magnetic field on the disk. It
turns out that the screw instability is related intimately to the
two parameters, i.e., the BH spin and the power-law index for the
variation of the magnetic field on the disk. In addition, the
correlation of the screw instability with CEBZMC is discussed. It
is shown that the state of CEBZMC always accompanies the screw
instability, and the region of the screw instability can be
determined by the criterion for the given values of the BH spin
and the power-law index. In \S 4 we discuss the effect of the
screw instability on magnetic extraction of energy from the
rotating BH. It is shown that the screw instability contributes
only a small fraction of magnetic extraction of energy from a
rotating BH. Finally, in \S 5, we summarize our main results.

Throughout this paper the geometric units $G = c = 1$ are used.

\section{DESCRIPTION OF THE MODEL OF CEBZMC}

Considering that the remote load in the BZ process and the load
disk in the MC process are connected with a rotating BH by open
and closed magnetic field lines, respectively, we studied a model
of CEBZMC (WXL, WMLY). The poloidal configuration of the magnetic
field is shown in Figure 1, where $r_{in} $ and $r_{out} $ are the
radii of the inner and outer boundaries of the MC region,
respectively. The angle $\theta _M $ indicates the angular
boundary between the open and closed field lines on the horizon,
and $\theta _L $ is the lower boundary angle for the closed field
lines. The angles $\theta _M $ and $\theta _L $ are connected with
$r_{out} $ and $r_{in} $ by the highest and the lowest closed
field lines, respectively. Throughout this paper $\theta _L =
0.45\pi $ is assumed in calculation. Following Blandford (1976),
we assume that the poloidal magnetic field $B_D^p $ on the disk
varies with the radial coordinate $\xi $ as a power law,

\begin{equation}
\label{eq1}
B_D^p \propto \xi ^{ - n},
\end{equation}

\noindent
Hereafter the subscript ``$D$'' is used to indicate the
quantities on the disk. The parameter $n$ is the power-law index
and $\xi \equiv r \mathord{\left/ {\vphantom {r {r_{ms} }}}
\right. \kern-\nulldelimiterspace} {r_{ms} }$ is the radial
coordinate on the disk, which is defined in terms of the radius
$r_{ms} \equiv M\chi _{ms}^2 $ of the marginally stable orbit
(Novikov {\&} Thorne 1973).


An equivalent circuit for the BZ and MC processes is used as shown in Figure
2, in which the symbols have the same meanings as given in WXL.


A stationary, axisymmetric magnetosphere anchored in a Kerr BH and its
surrounding disk can be described in Boyer-Lindquist coordinates, and the
concerned metric parameters are given as follows (MacDonald and Thorne 1982,
hereafter MT).

\begin{equation}
\label{eq2}
\left. {\begin{array}{l}
\Sigma ^2 = \left( {r^2 + a^2} \right)^2 - a^2\Delta \sin ^2\theta
,\quad\mbox{ }\rho ^2  = r^2 + a^2\cos ^2\theta ,\mbox{ } \\
\Delta  = r^2 + a^2 - 2Mr,\mbox{ }\quad \quad\quad\quad\quad\varpi= \left(
{\Sigma
\mathord{\left/ {\vphantom {\Sigma \rho }} \right.
\kern-\nulldelimiterspace} \rho } \right)\sin \theta . \\
\end{array}} \right\} ,
\end{equation}

\noindent
where $M$ and $a \equiv J \mathord{\left/ {\vphantom {J M}} \right.
\kern-\nulldelimiterspace} M$ are the mass and the specific angular momentum
of a Kerr BH, respectively. The current in each loop of the equivalent
circuit for the BZ and MC processes is expressed by

\begin{equation}
\label{eq3}
I_i = \frac{\Delta \varepsilon _{Hi} + \Delta \varepsilon _{Li} }{\Delta
Z_{Hi} + \Delta Z_{Li} } = \left( {{\Delta \Psi _i } \mathord{\left/
{\vphantom {{\Delta \Psi _i } {2\pi }}} \right. \kern-\nulldelimiterspace}
{2\pi }} \right)\frac{\Omega ^H - \Omega _i^F }{\Delta Z_{Hi} },
\end{equation}

\noindent
where the quantities in equation (\ref{eq3}) have the same meaning as given
in WXL.
In WMLY we derived the poloidal currents on the horizon as follows.

\begin{equation}
\label{eq4}
I_{H,BZ}^p \left( {a_ * ;\theta ,n} \right) = I_0 \frac{a_ * \left( {1 - k}
\right)}{2\csc ^2\theta - \left( {1 - q} \right)},
\quad
0 < \theta < \theta _M ,
\end{equation}

\begin{equation}
\label{eq5}
I_{H,MC}^p \left( {a_ * ;\theta ,n} \right) = I_0 \frac{a_ * \left( {1 -
\beta } \right)}{2\csc ^2\theta - \left( {1 - q} \right)},
\quad
\theta _M < \theta < \theta _L ,
\end{equation}

\begin{equation}
\label{eq6}
I_0 \equiv B_H^p M \approx B_4 \left( {M \mathord{\left/ {\vphantom {M {M_
\odot }}} \right. \kern-\nulldelimiterspace} {M_ \odot }} \right)\times
1.48\times 10^{10}A,
\end{equation}

\noindent where the subscripts ``BZ'' and ``MC'' represent the
quantities involving the BZ and MC processes, respectively, and
$B_4 $ is the strength of the magnetic field in units of
$10^4\mbox{ }G$. In equations (\ref{eq4}) and (\ref{eq5}) the
parameter $a_ * \equiv a \mathord{\left/ {\vphantom {a M}} \right.
\kern-\nulldelimiterspace} M$ is the BH spin, and the parameter
$q$ is defined as $q = \sqrt {1 - a_ * ^2 } $. The quantity $B_H^p
$ is the poloidal component of the magnetic field on the horizon.
Hereafter the subscript ``$H$'' is used to indicate the quantities
on the horizon.

The parameters $k$ and $\beta $ are the ratios of the angular velocity of
the open and closed magnetic field lines to that of the BH, respectively.
Usually, $k = 0.5$ is taken for the optimal BZ power, while $\beta $ depends
on the BH spin $a_ * $ and $\xi $. Since the closed field lines are frozen
in the disk, $\beta $ is expressed by

\begin{equation}
\label{eq7}
\beta \equiv {\Omega _D } \mathord{\left/ {\vphantom {{\Omega _D } {\Omega
_H }}} \right. \kern-\nulldelimiterspace} {\Omega _H } = \frac{2\left( {1 +
q} \right)}{a_ * }\left[ {\left( {\sqrt \xi \chi _{ms} } \right)^3 + a_ * }
\right]^{ - 1},
\end{equation}

\noindent
where $\Omega _H $ and $\Omega _D $ are the angular velocities of the BH and
the disk, respectively.

The poloidal current flowing on the disk is equal to that flowing on the
horizon in the same loop of Figure 2, i.e.,

\begin{equation}
\label{eq8}
I_{D,MC}^p \left( {a_ * ;\xi ,n} \right) = I_{H,MC}^p = I_0 \frac{a_ *
\left( {1 - \beta } \right)}{2\csc ^2\theta - \left( {1 - q} \right)},
\quad
1 < \xi < \xi _{out} ,
\end{equation}

\noindent
where $\xi $ is related to the angular coordinate $\theta $ on the horizon
by the mapping relation given in WMLY,

\begin{equation}
\label{eq9}
\cos \theta - \cos \theta _L = \int_1^\xi {\mbox{G}\left( {a_ * ;\xi ,n}
\right)d\xi } ,
\end{equation}

\noindent
where

\begin{equation}
\label{eq10}
\mbox{G}\left( {a_ * ;\xi ,n} \right) = \frac{\xi ^{1 - n}\chi _{ms}^2 \sqrt
{1 + a_ * ^2 \chi _{ms}^{ - 4} \xi ^{ - 2} + 2a_ * ^2 \chi _{ms}^{ - 6} \xi
^{ - 3}} }{2\sqrt {\left( {1 + a_ * ^2 \chi _{ms}^{ - 4} + 2a_ * ^2 \chi
_{ms}^{ - 6} } \right)\left( {1 - 2\chi _{ms}^{ - 2} \xi ^{ - 1} + a_ * ^2
\chi _{ms}^{ - 4} \xi ^{ - 2}} \right)} }.
\end{equation}

\noindent
The condition for CEBZMC is discussed in WMLY based on
the following assumptions:

(\ref{eq1}) The theory of a stationary, axisymmetric magnetosphere anchored
in the
BH and its surrounding disk formulated in MT is applicable not only to the
BZ process but also to the MC process. The magnetosphere is assumed to be
force-free outside the BH and the disk.

(\ref{eq2}) The disk is both stable and perfectly conducting, and the closed
magnetic field lines are frozen in the disk. The disk is thin and Keplerian,
lies in the equatorial plane of the BH with the inner boundary being at the
marginally stable orbit.

(\ref{eq3}) The magnetic field is assumed to be constant on the horizon, and
to vary
as a power law with the radial coordinate of the disk as given by equation
(\ref{eq1}).

(\ref{eq4}) The magnetic flux connecting a BH with its surrounding disk
takes
precedence over that connecting the BH with the remote load.

It is found that the state of CEBZMC will occur, provided that the BH spin
$a_ * $ and the power-law index $n$ are greater than some critical values.
In the following sections we adopt the above assumptions to discuss the
screw instability.

\section{CRITERIOR OF SCREW INSTABILITY IN CEBZMC}

\subsection{Derivation of Criterion }

According to the the Kruskal-Shafranov criterion, the screw
instability will occur, if the magnetic field line turns around
itself about once, i.e.,

\begin{equation}
\label{eq11}
{\left( {{2\pi \varpi _D } \mathord{\left/ {\vphantom {{2\pi \varpi _D } L}}
\right. \kern-\nulldelimiterspace} L} \right)B_D^p } \mathord{\left/
{\vphantom {{\left( {{2\pi \varpi _D } \mathord{\left/ {\vphantom {{2\pi
\varpi _D } L}} \right. \kern-\nulldelimiterspace} L} \right)B_D^p } {B_D^T
}}} \right. \kern-\nulldelimiterspace} {B_D^T } < 1,
\end{equation}

\noindent
where $B_D^p $ and $B_D^T $ are the poloidal and toroidal components of the
magnetic field on the disk, respectively, and $\varpi _D $ is the
cylindrical radius on the disk and reads

\begin{equation}
\label{eq12} \varpi _D = {\Sigma _D }/\rho_D= \xi M\chi _{ms}^2
\sqrt {1 + a_* ^2 \xi ^{ - 2}\chi _{ms}^{ - 4} + 2a_ * ^2 \xi ^{ -
3}\chi _{ms}^{ - 6} } .
\end{equation}

\noindent
In deriving equation (\ref{eq12}) the Kerr metric
parameters given in equation (\ref{eq2}) are used. The quantity
$L$ in equation (\ref{eq11}) is the poloidal length of the closed
field line connecting the BH with the disk. The criterion
(\ref{eq11}) can be rewritten as

\begin{equation}
\label{eq13}
{B_D^p } \mathord{\left/ {\vphantom {{B_D^p } {B_D^T }}} \right.
\kern-\nulldelimiterspace} {B_D^T } = \left( {{B_D^p } \mathord{\left/
{\vphantom {{B_D^p } {B_H^p }}} \right. \kern-\nulldelimiterspace} {B_H^p }}
\right)\left( {{B_H^p } \mathord{\left/ {\vphantom {{B_H^p } {B_D^T }}}
\right. \kern-\nulldelimiterspace} {B_D^T }} \right) < L \mathord{\left/
{\vphantom {L {\left( {2\pi \varpi _D } \right)}}} \right.
\kern-\nulldelimiterspace} {\left( {2\pi \varpi _D } \right)}.
\end{equation}

\noindent

Considering the continuum of magnetic flux between the two adjacent magnetic
surfaces connecting the BH and the disk, we have

\begin{equation}
\label{eq14}
B_H^p 2\pi \left( {\varpi \rho } \right)_{r = r_H } d\theta = - B_D^p 2\pi
\left( {{\varpi \rho } \mathord{\left/ {\vphantom {{\varpi \rho } {\sqrt
\Delta }}} \right. \kern-\nulldelimiterspace} {\sqrt \Delta }}
\right)_{\theta = \pi \mathord{\left/ {\vphantom {\pi 2}} \right.
\kern-\nulldelimiterspace} 2} dr.
\end{equation}

\noindent

Incorporating equation (\ref{eq14}) with the mapping relation in
WMLY, we express the ratio ${B_D^p } \mathord{\left/ {\vphantom
{{B_D^p } {B_H^p }}} \right. \kern-\nulldelimiterspace} {B_H^p }$
as

\noindent
\begin{equation}
\label{eq15}
\frac{B_D^p }{B_H^p } = \frac{2\left( {1 + q} \right)\sqrt {1 + a_ * ^2 \xi
^{ - 2}\chi _{ms}^{ - 4} - 2\xi ^{ - 1}\chi _{ms}^{ - 2} } }{\xi \chi
_{ms}^4 \sqrt {1 + a_ * ^2 \xi ^{ - 2}\chi _{ms}^{ - 4} + 2a_ * ^2 \xi ^{ -
3}\chi _{ms}^{ - 6} } }\mbox{G}\left( {a_ * ;\xi ,n} \right).
\end{equation}

\noindent

The ratio ${B_H^p } \mathord{\left/ {\vphantom {{B_H^p } {B_D^T }}} \right.
\kern-\nulldelimiterspace} {B_D^T }$ can be calculated by using equation
(\ref{eq8}), and the radial current density flowing on the disk can be
written as

\begin{equation}
\label{eq16}
j_D^p \left( {a_ * ;\xi ,n} \right) = {I_{D,MC}^p \left( {a_ * ;\xi ,n}
\right)} \mathord{\left/ {\vphantom {{I_{D,MC}^p \left( {a_ * ;\xi ,n}
\right)} {\left( {2\pi \varpi _D } \right)}}} \right.
\kern-\nulldelimiterspace} {\left( {2\pi \varpi _D } \right)}.
\end{equation}

\noindent

By using Ampere's law the toroidal magnetic field $\textbf{B}_D^T
$ on the disk is expressed by

\begin{equation}
\label{eq17} \textbf{{B}}_D^T = 4\pi \textbf{j}_D^p \times
\textbf{n},
\end{equation}

\noindent where $\textbf{n}$ is the unit vector normal to the
disk. Incorporating equations (\ref{eq8}), (\ref{eq12}),
(\ref{eq16}) and (\ref{eq17}), we have

\begin{equation}
\label{eq18}
\frac{B_H^p }{B_D^T } = \frac{\xi \chi _{ms}^2 \sqrt {1 + a_ * ^2 \xi ^{ -
2}\chi _{ms}^{ - 4} + 2a_ * ^2 \xi ^{ - 3}\chi _{ms}^{ - 6} } \left[ {2\csc
^2\theta - \left( {1 - q} \right)} \right]}{2a_ * \left( {1 - \beta }
\right)}.
\end{equation}

\noindent
Substituting equations (\ref{eq15}) and (\ref{eq18})
into the criterion (\ref{eq13}), we have

\begin{equation}
\label{eq19}
\left( {{2\pi \varpi _D } \mathord{\left/ {\vphantom {{2\pi \varpi _D } L}}
\right. \kern-\nulldelimiterspace} L} \right)F\left( {a_ * ;\xi ,n} \right)
< 1,
\end{equation}

\noindent
where

\begin{equation}
\label{eq20}
F\left( {a_ * ;\xi ,n} \right) = \frac{\xi ^{1 - n}\left( {1 + q}
\right)\left[ {2\csc ^2\theta - \left( {1 - q} \right)} \right]}{2a_ *
\left( {1 - \beta } \right)}\sqrt {\frac{1 + a_ * ^2 \chi _{ms}^{ - 4} \xi
^{ - 2} + 2a_ * ^2 \chi _{ms}^{ - 6} \xi ^{ - 3}}{1 + a_ * ^2 \chi _{ms}^{ -
4} + 2a_ * ^2 \chi _{ms}^{ - 6} }} .
\end{equation}

Since the poloidal length of the closed field line $L$ can be estimated
approximately as half of the circumference with diameter $D$, we have

\begin{equation}
\label{eq21}
L = AD = A\int_{r_H }^r {\left( {g_{rr} } \right)^{1 / 2}dr} ,
\end{equation}

\noindent
where $D$ is the proper distance from the horizon to the place of disk where
the closed field line penetrates, and $A \approx 0.5\pi $ is taken. The
metric parameter $g_{rr} $ is expressed as

\begin{equation}
\label{eq22}
g_{rr} = {\rho ^2} \mathord{\left/ {\vphantom {{\rho ^2} \Delta }} \right.
\kern-\nulldelimiterspace} \Delta = \frac{r^2}{r^2 + a^2 - 2Mr} = \frac{\xi
^2\chi _{ms}^4 }{\xi ^2\chi _{ms}^4 + a_ * ^2 - 2\xi \chi _{ms}^2 }.
\end{equation}

Incorporating equations (\ref{eq12}), (\ref{eq21}) and (\ref{eq22}), we have

\begin{equation}
\label{eq23}
L \mathord{\left/ {\vphantom {L {\left( {2\pi \varpi _D } \right)}}} \right.
\kern-\nulldelimiterspace} {\left( {2\pi \varpi _D } \right)} = \frac{A
\mathord{\left/ {\vphantom {A {2\pi }}} \right. \kern-\nulldelimiterspace}
{2\pi }}{\xi \sqrt {1 + a_ * ^2 \xi ^{ - 2}\chi _{ms}^{ - 4} + 2a_ * ^2 \xi
^{ - 3}\chi _{ms}^{ - 6} } }\int_{\xi _H }^\xi {\frac{d\xi }{\sqrt {1 + a_ *
^2 \xi ^{ - 2}\chi _{ms}^{ - 4} - 2\xi ^{ - 1}\chi _{ms}^{ - 2} } }} ,
\end{equation}

\noindent
where $\xi _H \equiv {r_H } \mathord{\left/ {\vphantom {{r_H } {r_{ms} }}}
\right. \kern-\nulldelimiterspace} {r_{ms} } = {\left( {1 + q} \right)}
\mathord{\left/ {\vphantom {{\left( {1 + q} \right)} {\chi _{ms}^2 }}}
\right. \kern-\nulldelimiterspace} {\chi _{ms}^2 }$. Therefore equations
(\ref{eq19}), (\ref{eq20}) and (\ref{eq23}) consist of the criterion of the
screw instability in our
model.

Inspecting the criterion (\ref{eq11}), we find that the screw
instability may occur, if the ratio of $B_D^p $ to $B_D^T $ is
small enough. We can give an analysis before calculation. In fact,
we find that $B_D^T $ is proportional to $\xi ^{ - 1}$
approximately by combining equations (\ref{eq6}), (\ref{eq8}),
(\ref{eq16}) and (\ref{eq17}). Thus the ratio ${B_D^p }
\mathord{\left/ {\vphantom {{B_D^p } {B_D^T }}} \right.
\kern-\nulldelimiterspace} {B_D^T }$ should be proportional to
$\xi ^{1 - n}$ by considering equation (\ref{eq1}), and it is
exactly the leading factor of the function $F\left( {a_ * ;\xi ,n}
\right)$. In addition, from equations (\ref{eq8}), (\ref{eq16})
and (\ref{eq17}) we find that the larger value of the BH spin
results in the larger values of $j_D^p $ and $B_D^T $. Therefore
we have the following conjectures on the screw instability.

(\ref{eq1}) The screw instability would occur at some place far away from
the inner
edge of the disk.

(\ref{eq2}) The screw instability would occur, provided that the BH spin $a_
* $ and
the power-law index $n$ are greater than some critical values.

The above conjectures are verified by the calculations on the
criterion (\ref{eq19}) as shown in Figure 3.


In Figure 3a $\left( {{2\pi \varpi _D } \mathord{\left/ {\vphantom
{{2\pi \varpi _D } L}} \right. \kern-\nulldelimiterspace} L}
\right)F\left( {a_ * ;\xi ,n} \right) < 1$ holds for $0.8903 < a_*
<1 $ with $n = 4$, and in Figure 3b it holds for $n > 3.34$ with
$a_* = 0.998$. These results imply that the screw instability
occurs, if the BH spin and the power-law index are greater than
some critical values.

\subsection{Screw Instability and CEBZMC}

In WMLY we argued that the state of CEBZMC will occur, if the parameters $a_
* $ and $n$ are greater than some critical values. So it is tempting for us
to discuss the correlation of the screw instability with CEBZMC. Based on
WMLY and the criterion (\ref{eq19}) we have the critical line for CEBZMC and
the
critical lines for the screw instability in the parameter space consisting
of $a_ * $ and $n$ as shown in Figure 4.


From Figure 4 we obtain the following results.

(\ref{eq1}) There is very little difference among the critical lines for the
screw
instability with different values of the factor $A$, and these critical
lines can not be distinguished in Figure 4a, and are very near to the
critical line for CEBZMC (thick solid line) as shown in Figure 4b.
Therefore, the screw instability is rather insensitive to the values of the
factor $A$.

(\ref{eq2}) It is shown in Figure 4b that the critical line for CEBZMC
corresponds
to the larger values of $a_ * $ and $n$ than the critical lines for the
screw instability. It implies that the state of CEBZMC always accompanies
the screw instability.

\subsection{ Disk Region for Screw Instability}

From the above discussion we conclude that the minimum radial coordinate
$\xi _{screw} $ for the screw instability can be determined by the following
equation,

\begin{equation}
\label{eq24}
\left( {{2\pi \varpi _D } \mathord{\left/ {\vphantom {{2\pi \varpi _D } L}}
\right. \kern-\nulldelimiterspace} L} \right)F\left( {a_ * ;\xi _{screw} ,n}
\right) = 1.
\end{equation}

Equation (\ref{eq24}) implies that the disk region for the screw instability
is

\begin{equation}
\label{eq25}
\xi _{screw} < \xi < \infty .
\end{equation}

By using the mapping relation (\ref{eq9}) and equation (\ref{eq25}), we
obtain the
corresponding angular region on the horizon as follows,

\begin{equation}
\label{eq26}
\theta _M < \theta < \theta _{screw} .
\end{equation}

By using equations (\ref{eq19}), (\ref{eq20}) and (\ref{eq23}) we have the
contours of constant
values of $\xi _{screw} $ for the screw instability in the parameter space
consisting of $a_ * $ and $n$ as shown in Figure 5.


From Figure 5 we find that the screw instability could even take place in
the inner region of the disk with a small value of $\xi _{screw} $, provided
that $a_ * $ and $n$ are great enough. Since the large values of $a_ * $ and
$n$ in the MC process are supported by the recent \textit{XMM-Newton}
observation for producing
a very steep emissivity in the inner region of the disk (WLM, WMLY), these
values imply not only CEBZMC, but also the screw instability.

\section{ SCREW INSTABILITY AND ENERGY EXTRACTION FROM A ROTATING
BH}

Gruzinov (1999) argued that the screw instability can be regarded as a new
mechanism of extracting rotational energy from a BH. His argument is based
on a model with a bunch of closed magnetic field lines connecting a Kerr BH
with a disk. Compared with Gruzinov's model, the magnetic field in our model
is axisymmetric, and the region of the screw instability is determined by
equation (\ref{eq25}) rather than is given at random.

Considering the energy released due to the screw instability, the total
power of the magnetic extraction from the BH consists of the following three
parts, i.e.,

\begin{equation}
\label{eq27}
P_{total} = P_{BZ} + P_{MC} + \bar {P}_{screw} ,
\end{equation}

\noindent
where $P_{BZ} $, $P_{MC} $ and $\bar {P}_{screw} $ are the BZ power, the MC
power and the average power due to the screw instability, respectively.
Incorporating the region for screw instability given by equation
(\ref{eq26}) with
the expressions for the BZ and MC powers given in WXL, we have

\begin{equation}
\label{eq28}
{P_{BZ} } \mathord{\left/ {\vphantom {{P_{BZ} } {P_0 }}} \right.
\kern-\nulldelimiterspace} {P_0 } = 2a_ * ^2 \int_0^{\theta _M }
{\frac{k\left( {1 - k} \right)\sin ^3\theta d\theta }{2 - \left( {1 - q}
\right)\sin ^2\theta }} ,
\end{equation}

\begin{equation}
\label{eq29}
{P_{MC} } \mathord{\left/ {\vphantom {{P_{MC} } {P_0 }}} \right.
\kern-\nulldelimiterspace} {P_0 } = 2a_ * ^2 \int_{\theta _{screw} }^{\theta
_L } {\frac{\beta \left( {1 - \beta } \right)\sin ^3\theta d\theta }{2 -
\left( {1 - q} \right)\sin ^2\theta }} ,
\end{equation}

\noindent
and

\begin{equation}
\label{eq30}
{\bar {P}_{screw} } \mathord{\left/ {\vphantom {{\bar {P}_{screw} } {P_0 }}}
\right. \kern-\nulldelimiterspace} {P_0 } = 2a_ * ^2 \int_{\theta _M
}^{\theta _{screw} } {\frac{\beta \left( {1 - \beta } \right)\sin ^3\theta
d\theta }{2 - \left( {1 - q} \right)\sin ^2\theta }} .
\end{equation}
\noindent
where $P_0 = {(B_H^p)^2}M^2\approx
B_4^2{(M/M_\odot)^2}\times6.59\times10^{28}erg\cdot s^{-1}.$

By using equations (\ref{eq28})---(\ref{eq30}) we obtain the
curves of ${P_{BZ} } \mathord{\left/ {\vphantom {{P_{BZ} } {P_0
}}} \right. \kern-\nulldelimiterspace} {P_0 }$, ${P_{MC} }
\mathord{\left/ {\vphantom {{P_{MC} } {P_0 }}} \right.
\kern-\nulldelimiterspace} {P_0 }$ and ${\bar {P}_{screw} }
\mathord{\left/ {\vphantom {{\bar {P}_{screw} } {P_0 }}} \right.
\kern-\nulldelimiterspace} {P_0 }$ varying with the parameters $a_
* $ and $n$ as shown in Figures 6 and 7, respectively. Defining
the ratio of $\bar {P}_{screw} $ to the total power $P_{total} $
as $\eta _P \equiv {\bar {P}_{screw} } \mathord{\left/ {\vphantom
{{\bar {P}_{screw} } {P_{total} }}} \right.
\kern-\nulldelimiterspace} {P_{total} }$, we have the curves of
$\eta _P $ varying with $a_ * $ and $n$ as shown in Figure 8. In
the following calculation $A = 0.5 \pi$ is assumed.



\noindent

From Figures 6, 7 and 8 we obtain the following results:

(\ref{eq1}) For the given value of $n$, $P_{BZ} $, $P_{MC} $ and $\bar
{P}_{screw} $
all vary non-monotonically with the BH spin $a_ * $, attaining the maxima as
$a_ * $ approaching unity.

(\ref{eq2}) For the given value of $a_ * $, $P_{BZ} $ increases
monotonically, $P_{MC} $ decreases monotonically , while $\bar
{P}_{screw} $ varies non-monotonically with the power-law index
$n$.

(\ref{eq3}) The average power $\bar {P}_{screw} $ is always less than
$P_{MC} $, and
it is less than $P_{BZ} $ in most cases, except that the power-law index is
small as shown in Figure 7a.

(\ref{eq4}) The ratio of power, $\eta _P $, increases monotonically with $a_
* $,
while it varies non-monotonically with the power-law index $n$, attaining
its maximum $\eta _P \approx 0.084$ at $n \approx 5.25$ with $a_ * \approx
0.997$.

Therefore we conclude that the screw instability contributes only a small
fraction of magnetic extraction of energy from a rotating BH.

\section{ SUMMARY}

In this paper the screw instability of the magnetic field
connecting a rotating BH with its surrounding disk is discussed in
the simplified model of CEBZMC. In our model the configuration of
the poloidal component of the magnetic field is assumed, which
remains constant on the horizon, and varies with the radial
coordinate on the disk according to a power law. By using an
equivalent circuit for CZBZMC, we can calculate the poloidal
current on the horizon and the disk, and then the toroidal
component of the magnetic field on the disk is derived by using
Ampere's law. Therefore, the criterion of the screw instability in
our model is derived based on the Kruskal-Shafranov criterion.

Although the discussion is carried out under some simplified assumptions,
several interesting results for the screw instability have been obtained.

(\ref{eq1}) It is shown that the BH spin $a_ * $ and the power-law index $n$
are
involved in the criterion, and the screw instability will occur if the two
parameters are greater than some critical values. This result means that the
screw instability is apt to take place for a disk containing a fast-spinning
BH, and has the magnetic field concentrating in the inner region.

(\ref{eq2}) We find that the disk region for the screw instability
can be also determined by $a_ * $ and $ n$ by virtue of the
criterion, and the screw instability is apt to take place outside
some critical radius on the disk.

(\ref{eq3}) We prove that the state of CEBZMC always accompanies
the screw instability as shown in the parameter space consisting
of $a_ * $ and $ n$.

(\ref{eq4}) As an energy mechanism the effect of the screw
instability on the magnetic extraction is discussed. It turns out
that the contribution of the screw instability is not significant
in our model of CEBZMC.

In this paper we fail to discuss the screw instability of the magnetic field
in the BZ process, since we know very few about the remote astrophysical
load. Considering stringent upper bound to the BZ power due to the screw
instability in the BZ process (Li 2000a), we should give an overall
evaluation on the effects of the screw instability on energy extraction from
a rotating BH in the future work.

\acknowledgments
{\bf Acknowledgements:}
This work is supported by the National Natural
Science Foundation of China under Grant Numbers 10173004 and
10121503.

\clearpage

\begin{figure}
\begin{center}
\epsscale{0.5}
\plotone {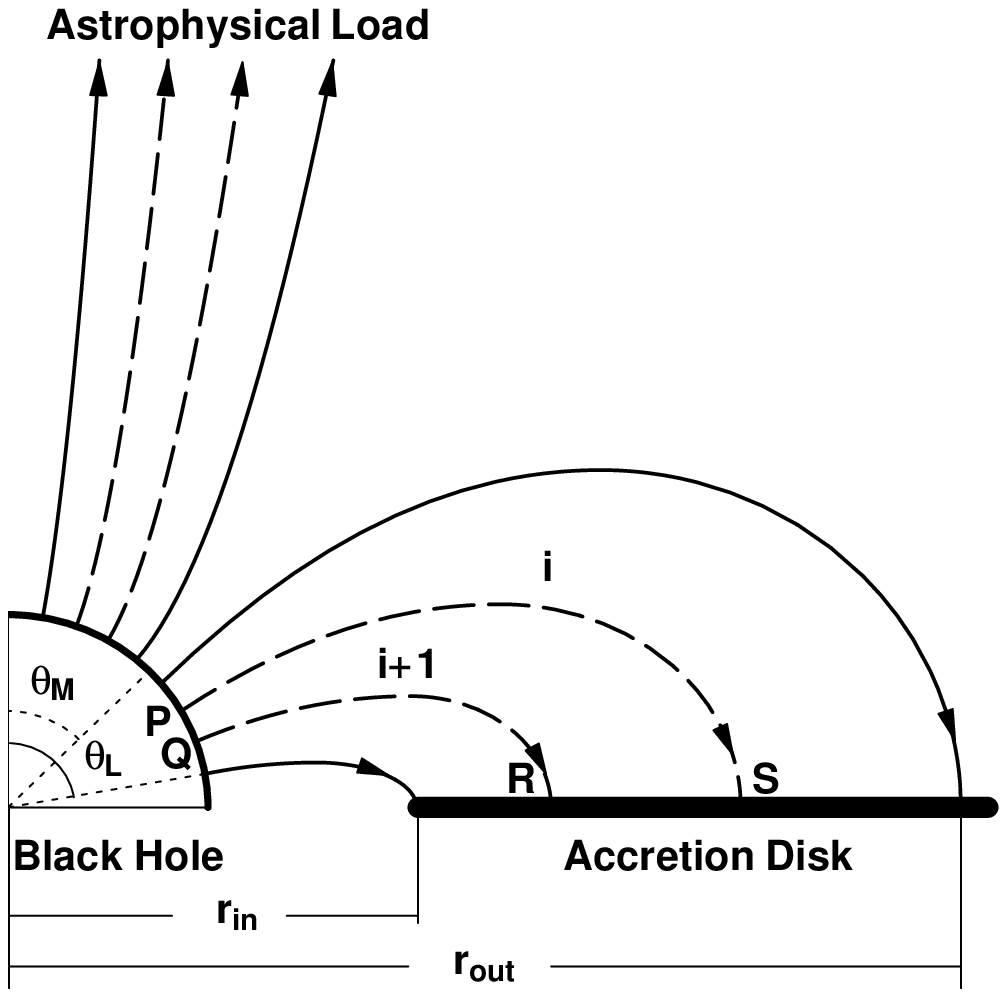}
\end{center}
\caption{ Poloidal magnetic field connecting a rotating BH with a
remote astrophysical load and the surrounding disk.}\label{fig1}

\end{figure}

\clearpage

\begin{figure}
\begin{center}
\epsscale{0.8}
\plotone{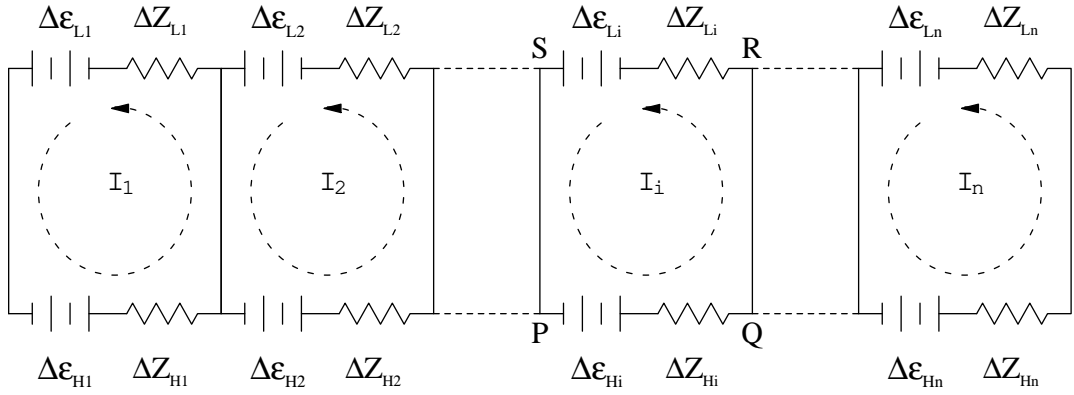}
\end{center}
\caption{Equivalent circuit for a model of CEBZMC.}\label{fig2}
\end{figure}

\clearpage

\begin{center}
\begin{figure}
\epsscale{0.4}
\begin{center}
  \plotone{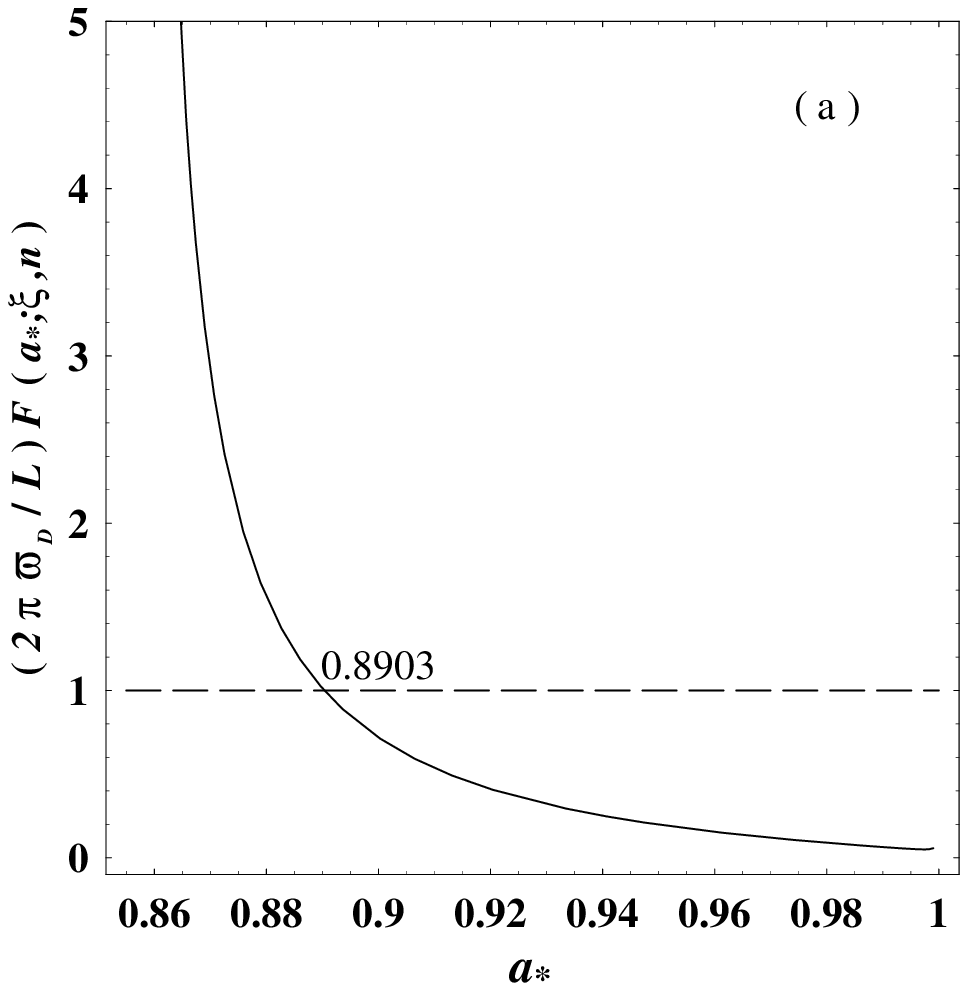}
\end{center}
\begin{center}
\epsscale{0.4}
  \plotone{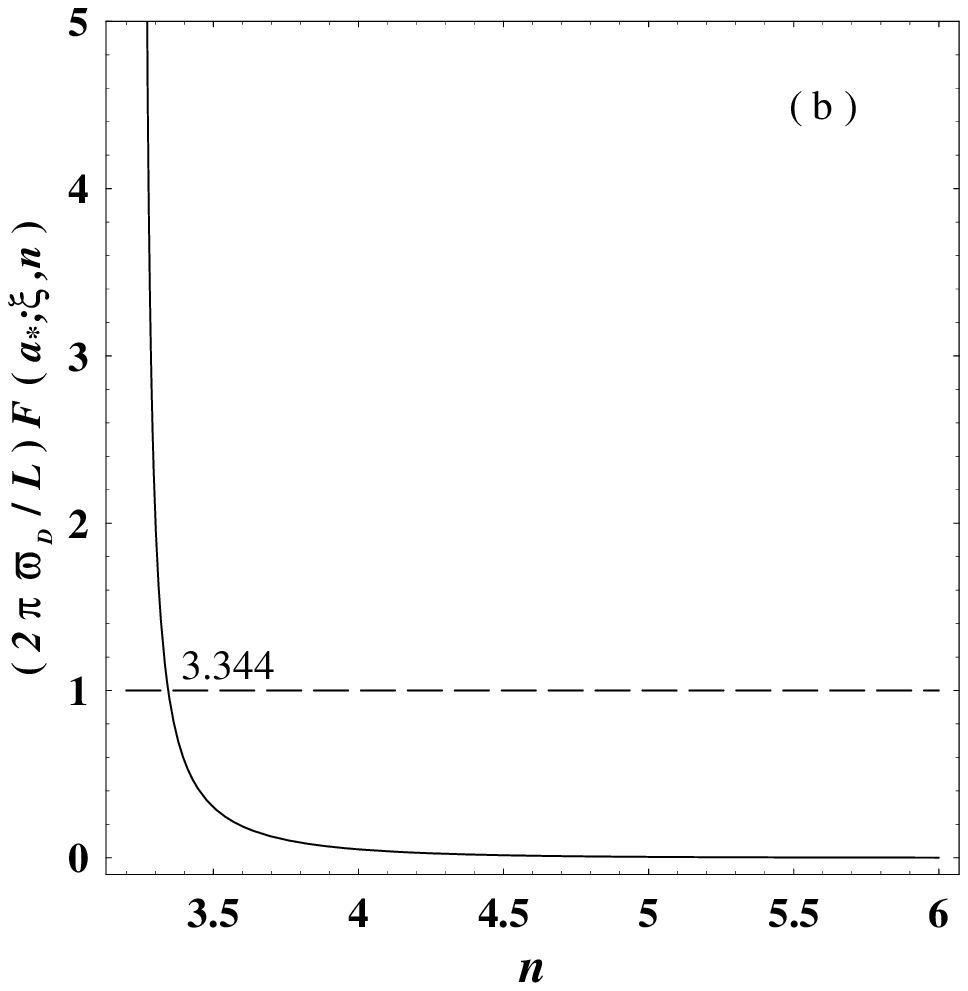}
\end{center}
\caption{Value of $( 2\pi \varpi_D / L )F( a_* ;\xi ,n )$ at $\xi
= 5$ (a) varying with $a_ * $ for $n = 4$ (b) varying with $n$ for
$a_* = 0.998$.}\label{fig3}
\end{figure}
\end{center}

\clearpage
\begin{figure}
\begin{center}
\epsscale{0.35}
\plotone{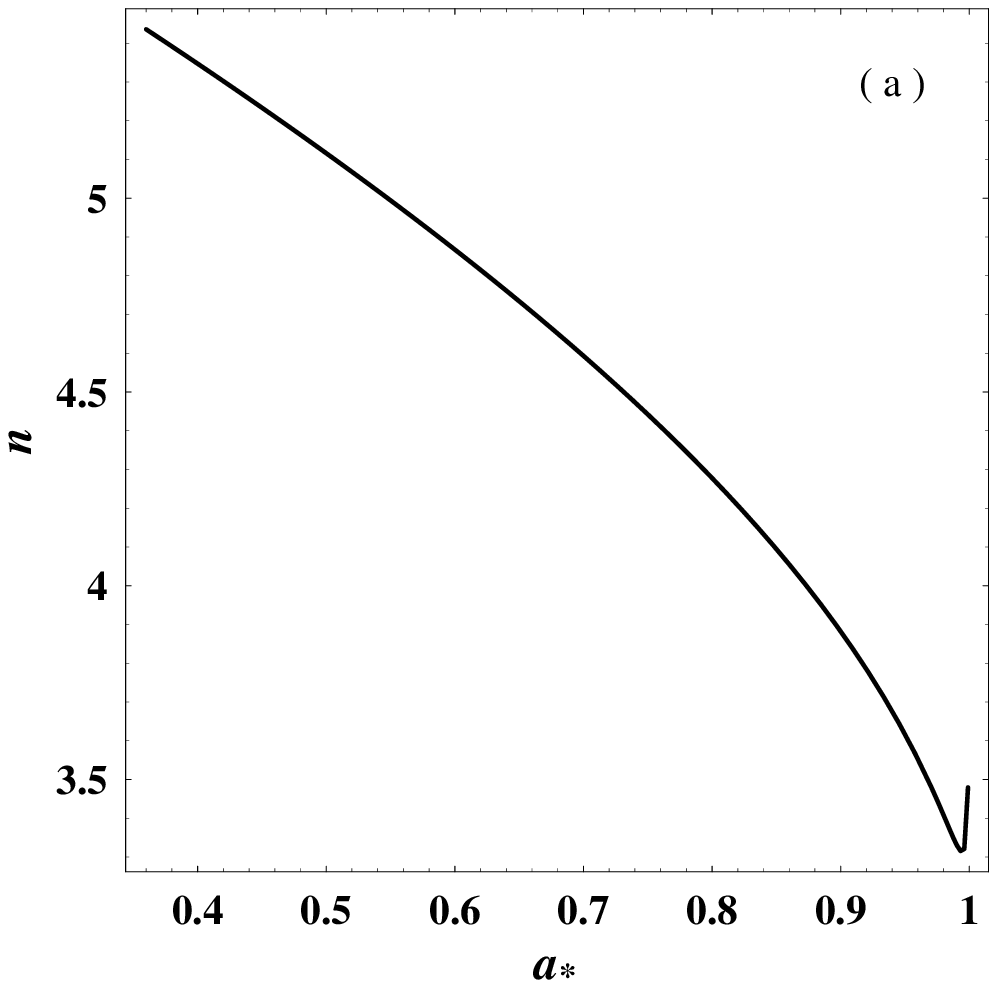}
\end{center}
\begin{center}
\epsscale{0.37} \plotone{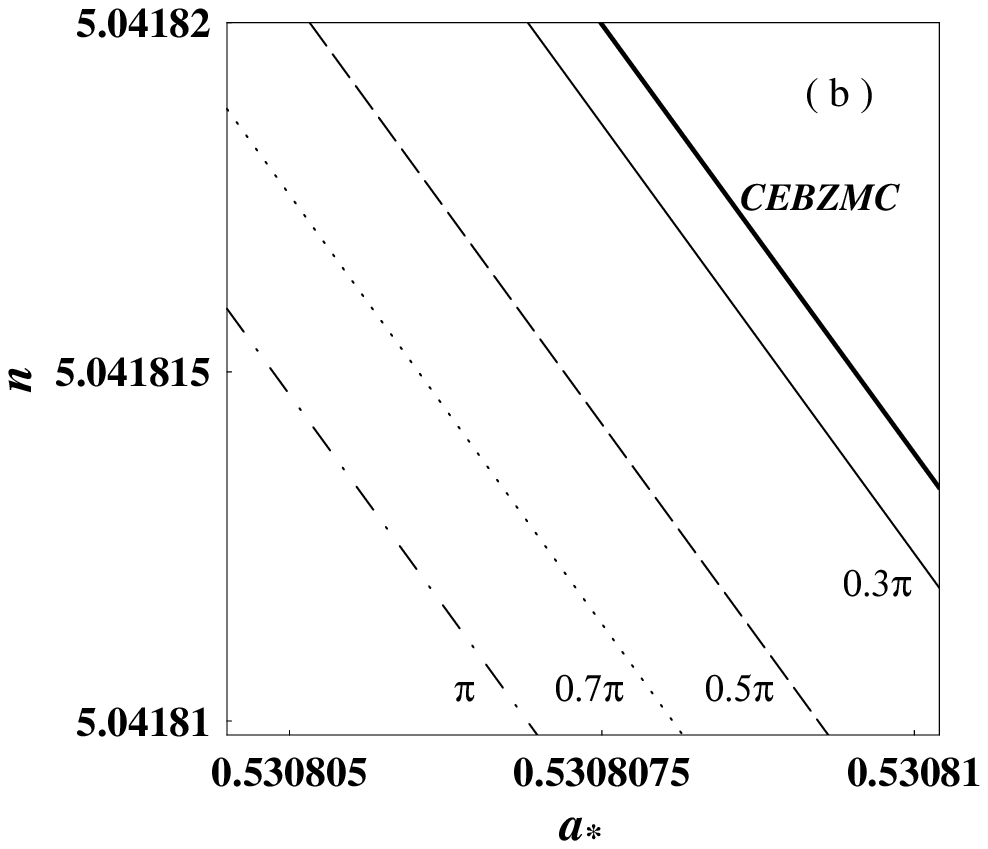}
\end{center}

\caption {Critical line for CEBZMC (thick solid line) and critical
lines for screw instability with different values of factor $ A =
0.3\pi $, 0.5$\pi $, 0.7$\pi $,$\pi $ in solid, dashed, dotted and
dot-dashed lines, respectively.}\label{fig4}
\end{figure}

\clearpage

\begin{figure}
\begin{center}
\epsscale{0.5}
\plotone{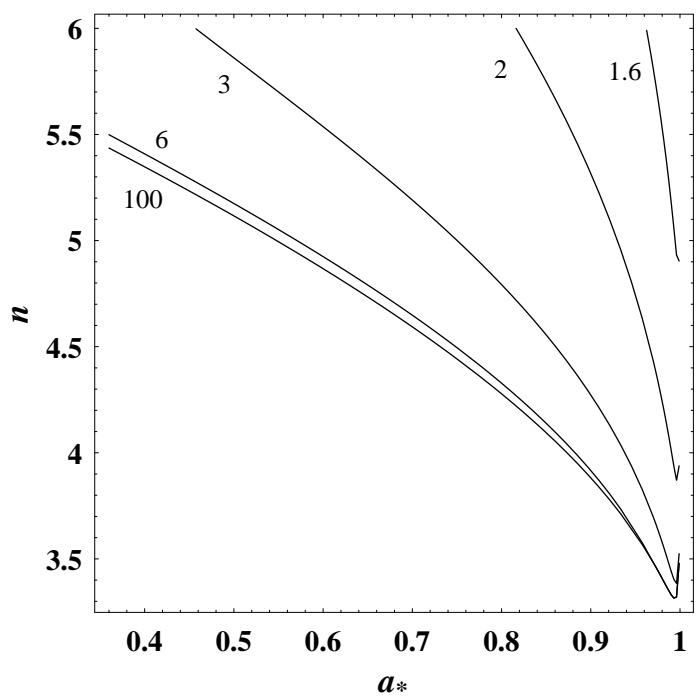}
\caption{Contours of constant values of $\xi _{screw} $ for the
screw instability with $A = 0.5\pi $.}\label{fig5}
\end{center}
\end{figure}

\clearpage

\begin{center}
\begin{figure}
\begin{center}
\epsscale{0.35}
\plotone{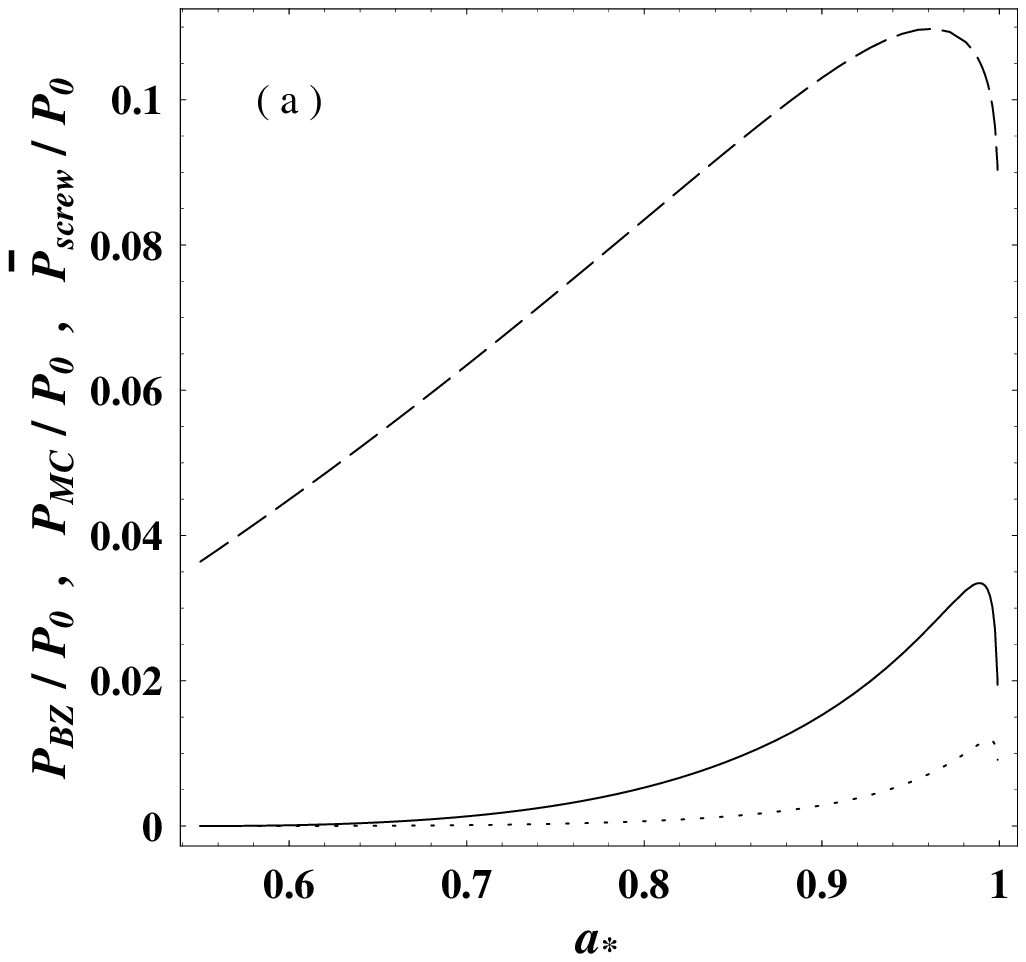}
\end{center}
\begin{center}
\epsscale{0.35}
\plotone{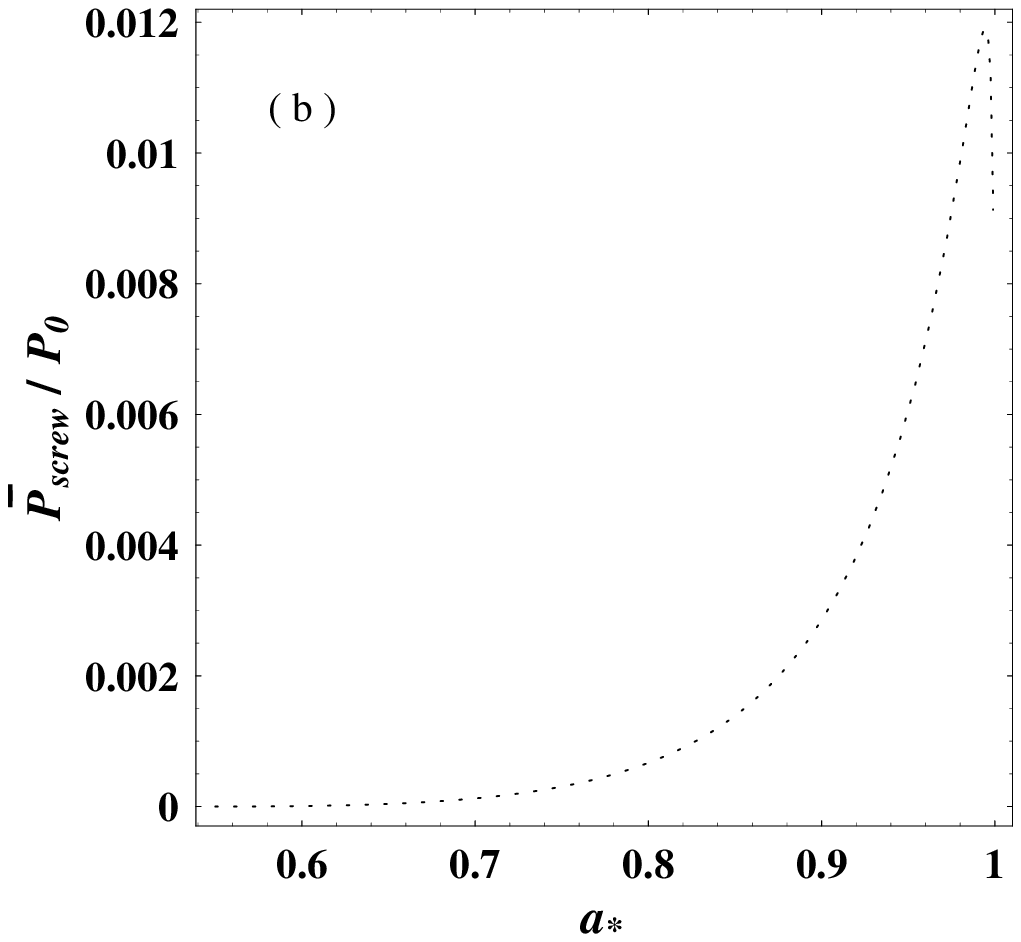}
\end{center}
\caption {(a) Curves of ${P_{BZ}}/ {P_0 }$(solid
line),${P_{MC}}/{P_0} $(dashed line), and ${\bar {P}_{screw}}/{P_0
}$ (dotted line), and (b) curve of ${\bar {P}_{screw}}/{P_0}$
(dotted line) varying with $ a_* $ for $n = 5$.}\label{fig6}
\end{figure}
\end{center}

\clearpage
\begin{center}
\begin{figure}
\begin{center}
\epsscale{0.35}
\plotone{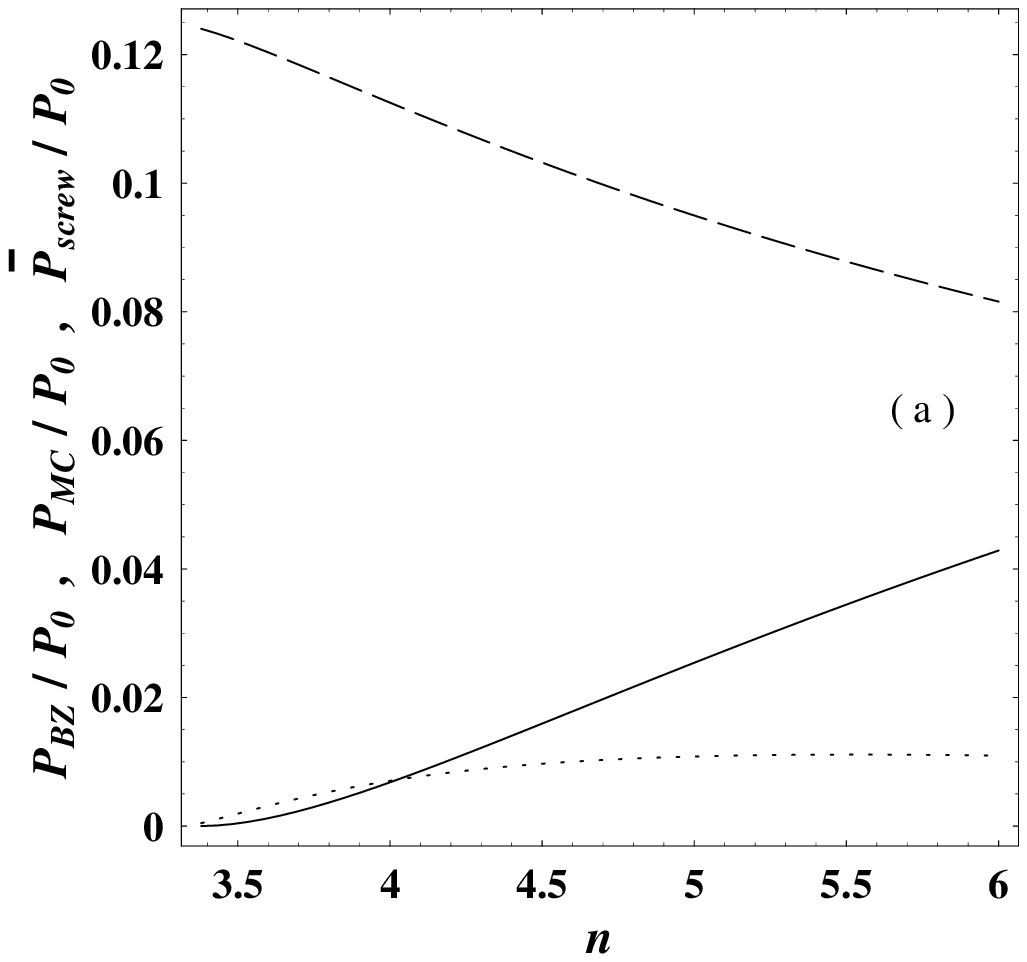}
\end{center}
\begin{center}
\epsscale{0.35}
\plotone{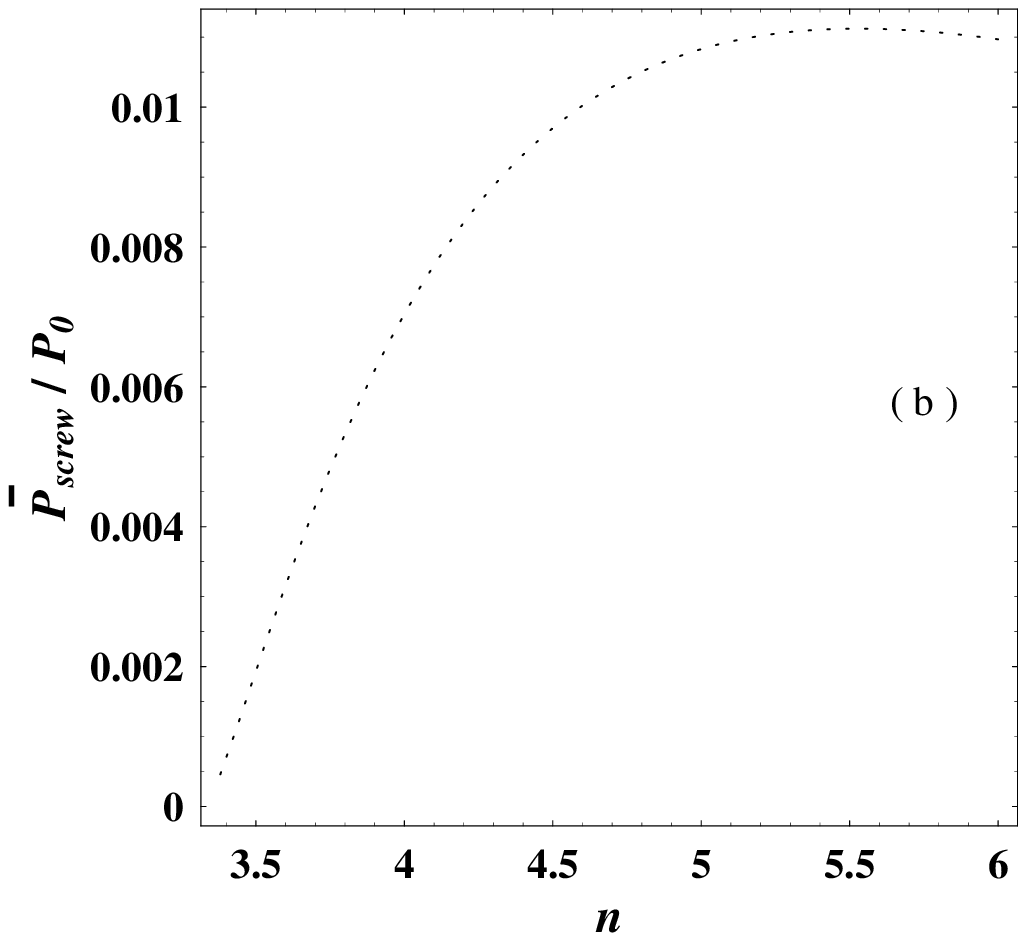}
\end{center}
\caption{(a) Curves of ${P_{BZ}}/ {P_0 }$(solid
line),${P_{MC}}/{P_0} $(dashed line), and ${\bar {P}_{screw}}/{P_0
}$ (dotted line), and (b) curve of ${\bar {P}_{screw}}/{P_0}$
(dotted line) varying with $ n $ for $a_* =0.998$.}\label{fig7}
\end{figure}
\end{center}

\clearpage

\begin{center}
\begin{figure}
\begin{center}
\epsscale{0.35}
\plotone{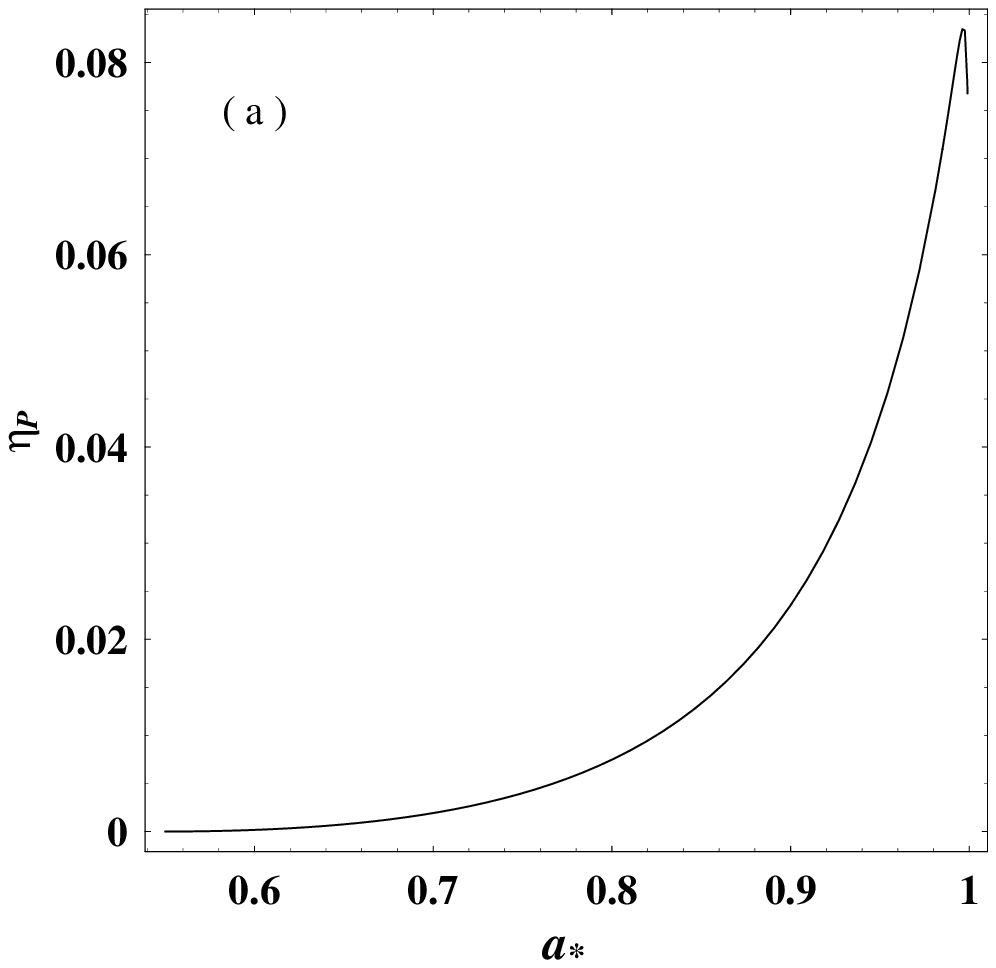}
\end{center}
\begin{center}
\epsscale{0.35}
\plotone{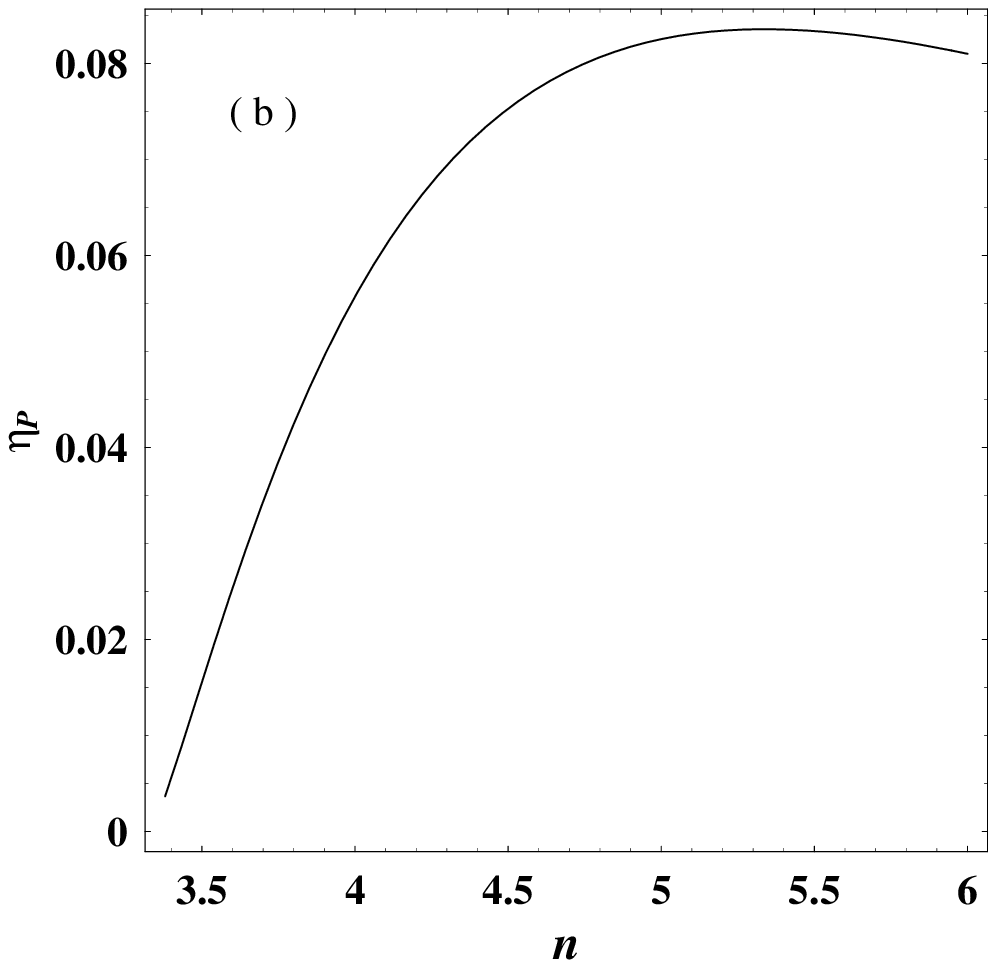}
\end{center}
\caption{Ratio of $\bar {P}_{screw} $ to total power $P_{total} $
(a) varying with $a_ * $ for $n = 5$, (b) varying with $n$ for $a_
* = 0.998$.}\label{fig8}
\end{figure}
\end{center}

\end{document}